\documentclass[letterpaper]{article} 
\usepackage{aaai2026}  
\usepackage{times}  
\usepackage{helvet}  
\usepackage{courier}  
\usepackage[hyphens]{url}  
\usepackage{graphicx} 
\usepackage{natbib}  
\usepackage{caption} 
\usepackage{multirow}
\usepackage{fontspec}
\usepackage{array}
\newcolumntype{C}[1]{>{\centering\arraybackslash}p{#1}}
\newcolumntype{L}[1]{>{\raggedright\arraybackslash}p{#1}}

\usepackage{amsmath}
\usepackage{amssymb}
\usepackage{booktabs}
\usepackage{algorithm}
\usepackage{algorithmic}
\usepackage[most]{tcolorbox}
\usepackage{listings}
\usepackage{xcolor}

\definecolor{promptblue}{HTML}{EAF4F8}
\definecolor{promptborder}{HTML}{D6E7EE}
\newcommand{\metricimp}[3]{%
  \rule[-0.7ex]{0pt}{3.0ex}%
  \begingroup
  \boldmath
  \textbf{\ensuremath{#1_{\scriptscriptstyle \pm #2}}}%
  \endgroup
  \makebox[0pt][l]{\,{\color{darkred}\scriptsize\bfseries\ensuremath{#3}}}%
}

\lstdefinestyle{promptstyle}{
    basicstyle=\ttfamily\scriptsize,
    breaklines=true,
    breakatwhitespace=false,
    columns=fullflexible,
    keepspaces=true,
    showstringspaces=false,
    frame=none,
    upquote=true
}

\newtcblisting{promptblock}{
    enhanced,
    breakable,
    listing only,
    colback=promptblue,
    colframe=promptborder,
    boxrule=0.4pt,
    arc=0pt,
    left=5pt,
    right=5pt,
    top=5pt,
    bottom=5pt,
    listing options={style=promptstyle}
}

\usepackage[table]{xcolor}
\usepackage{makecell}

\definecolor{ourgray}{HTML}{EFEFEF}
\definecolor{darkred}{HTML}{8B0000}
\ifdefined\pdfpagewidth
\fi
\ifdefined
\fi

\nocopyright
\title{Dynamic Trust-Aware Sparse Communication Topology for LLM-Based Multi-Agent Consensus}

\author{
    Wanshuang Gou\textsuperscript{\rm 1},
    Zihan Liu\textsuperscript{\rm 1}
}
\affiliations{
    \textsuperscript{\rm 1} Chengdu University \\
}

\begin{document}
\maketitle

\begin{abstract}
Large language model–driven multi-agent systems enhance the reliability of complex reasoning tasks through multi-round deliberation, role specialization, and cross-validation. However, existing multi-agent debate and collaboration frameworks typically adopt fully connected communication, causing the number of messages, token costs, and end-to-end latency to grow approximately quadratically with the number of agents; although fixed sparse topologies reduce overhead, they cannot adapt communication relationships to different task instances or intermediate reasoning states, making them prone either to preserving low-value interactions or to losing critical error-correction information. To address this problem, this paper proposes DySCo (Dynamic Sparse Consensus), a dynamic trust-aware sparse consensus mechanism. In each round of reasoning, DySCo estimates the value of communication edges based on agent reliability, answer divergence, and task relevance, and selects a small number of high-value edges for message exchange under budget constraints; it then aggregates the answers of different agents through dynamic trust weights and terminates the discussion early once consensus stabilizes. This mechanism replaces universal broadcasting with on-demand communication, thereby reducing communication overhead while preserving essential cross-validation information. We further present analyses of communication complexity and consensus stability, and evaluate the performance of DySCo on mathematical reasoning, logical reasoning, and factual question-answering tasks.
\end{abstract}

\section{Introduction}
\label{sec:introduction}
In recent years, LLM-based multi-agent systems have emerged as an important paradigm for enhancing the capacity to solve complex tasks. Representative approaches include Multi-Agent Debate (MAD)\citep{du2023improving}, ChatDev\citep{qian2024chatdev}, AutoGen\citep{wu2023autogen}, and MetaGPT\citep{hong2024metagpt}. These systems typically organize multiple LLM instances into agents endowed with distinct roles, perspectives, or tool-use capabilities, enabling collaborative reasoning or task planning through the exchange of natural-language messages. Compared with a single LLM, multi-agent mechanisms can mitigate answer instability in complex reasoning by drawing upon diverse perspectives.

\begin{figure}[t]
    \centering
    \begin{minipage}{0.48\linewidth}
        \centering
        \includegraphics[width=\linewidth]{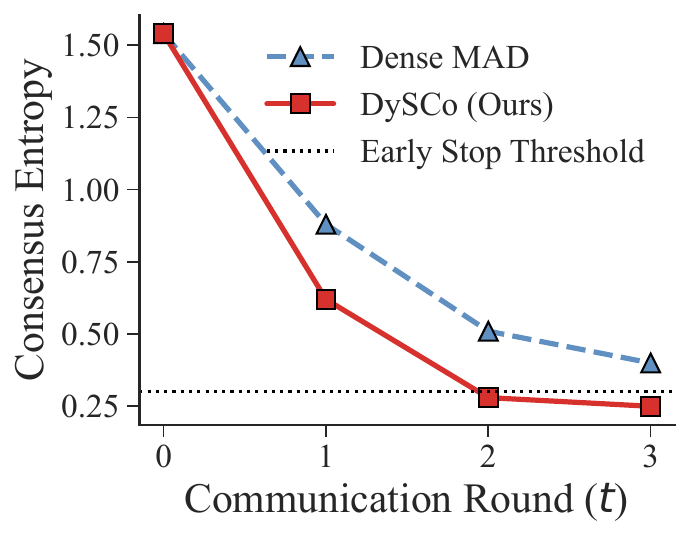}
        \centerline{(a)}
    \end{minipage}
    \hfill
    \begin{minipage}{0.48\linewidth}
        \centering
        \includegraphics[width=\linewidth]{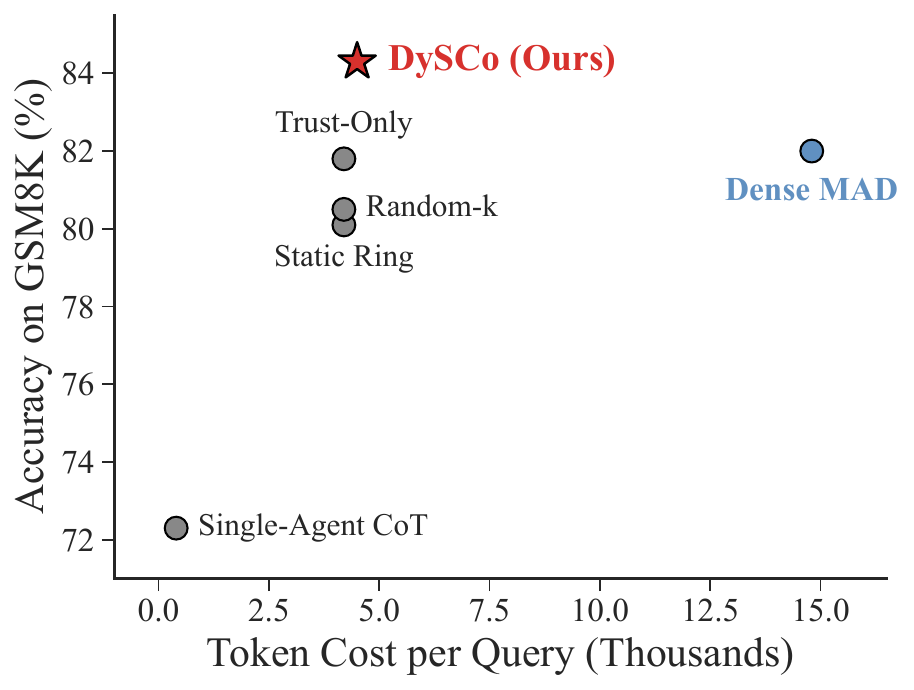}
        \centerline{(b)}
    \end{minipage}
    \caption{A schematic illustration of the core motivation and overall effect of DySCo. The left panel depicts the process of consensus convergence under different communication mechanisms, where DySCo more rapidly reduces consensus entropy among agents through dynamic sparse communication and can terminate early once a stable threshold is reached. The right panel illustrates the trade-off between accuracy and communication cost: compared with fully connected multi-agent debate, DySCo achieves superior reasoning performance with lower token costs, demonstrating the advantage of dynamic sparse communication in balancing efficiency and quality.}
    \label{fig:intro-summary}
\end{figure}

However, multi-agent collaboration is often accompanied by substantial communication overhead. For a system comprising $n$ agents and running for $R$ rounds of interaction, if fully connected communication is adopted in each round, the number of messages is $O(Rn^2)$. Since each message often contains intermediate reasoning or answer explanations, this overhead is directly translated into token cost and end-to-end latency. As the number of agents scales, fully connected communication can readily become a system bottleneck and may further amplify the propagation of flawed reasoning among agents.

Existing work has demonstrated that communication topology is a crucial factor shaping the efficiency and effectiveness of LLM-based multi-agent debate. \citet{li2024sparse} systematically investigated a range of sparse communication topologies and found that sparse communication can preserve, or even enhance, reasoning performance while reducing message exchange and computational cost. This indicates that multi-agent systems do not always require fully connected interaction, and that appropriately constraining the scope of communication can diminish the propagation of redundant information. However, existing methods primarily rely on fixed sparse topologies, such as ring graphs, star graphs, random graphs, or predefined graph structures. Such methods address the excessive cost of fully connected communication, but they do not resolve the mismatch between communication relationships and reasoning states. During multi-round reasoning, agents’ reliability, confidence, answer divergence, and task relevance continually evolve; a fixed topology may preserve low-value messages while also missing critical error-correction information. Therefore, how to dynamically select communication edges according to the current reasoning state constitutes a central problem in building an efficient consensus mechanism for LLM-based multi-agent systems.

To address the aforementioned problem, this paper proposes DySCo (Dynamic Sparse Consensus), a dynamic trust-aware sparse consensus mechanism. Rather than relying on a fixed communication graph, DySCo dynamically assesses the value of communication edges in each round of reasoning according to agent reliability, answer divergence, confidence, and task relevance, and selects a small number of high-value neighbors for information exchange under budget constraints. After receiving feedback, agents update their answers by integrating local reasoning with neighboring opinions, while the system forms a final consensus through trust-weighted aggregation and terminates early once consensus becomes stable. In this way, DySCo replaces fully connected broadcasting with dynamic sparse communication, preserving critical error-correction information while reducing token consumption and latency.

The contributions of this paper are as follows:
    \begin{itemize}
        \item We propose a dynamic sparse communication framework for LLM-based multi-agent reasoning, extending communication topology from a fixed structure into an adaptive decision variable driven by task states.
        \item We design a trust-aware edge selection mechanism that jointly considers historical reliability, current confidence, answer complementarity, and task dependencies, selecting high-value communication edges under a given token budget.
        \item We provide analyses of communication complexity and abstract consensus stability, showing that when each agent connects to at most $k$ neighbors in each round, the communication complexity can be reduced from $O(Rn^2)$ to $O(Rnk)$.
        \item We establish a reproducible experimental protocol for comparing fully connected communication, multiple static sparse topologies, and DySCo in terms of reasoning accuracy, token cost, latency, and consensus stability.
    \end{itemize}

\begin{figure*}[!h]
    \centering
    \includegraphics[width=\textwidth]{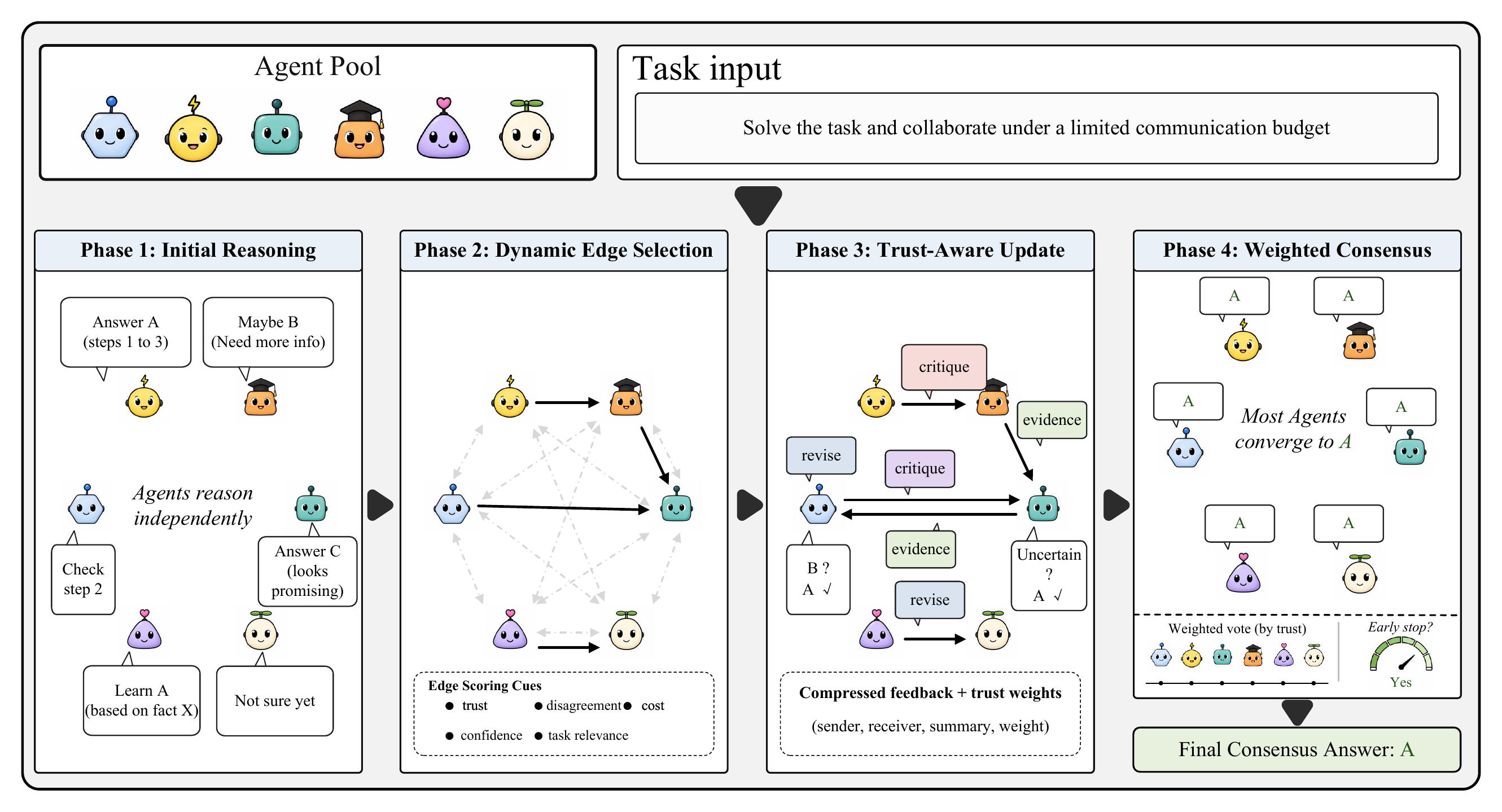}
    \caption{Overview of the DySCo framework. Given a task input and an agent pool, each LLM agent first reasons independently and generates an initial answer. Subsequently, DySCo evaluates the value of candidate communication edges according to historical trust, current confidence, answer divergence, task relevance, and communication cost, selecting a small number of high-value edges under budget constraints. Agents exchange compressed feedback only through the selected edges and update their own answers by incorporating trust weights. Finally, the system aggregates the outputs of all agents through trust-weighted voting and terminates early once consensus becomes stable.}
    \label{fig:qualitative_rollout}
\end{figure*}

\section{Related Work}

\subsection{LLM-Based Multi-Agent Collaboration and Debate}
LLM-based multi-agent systems enhance complex task-solving capabilities by organizing multiple model instances for role specialization, cross-agent discussion, or collaborative task execution. Unlike single-model sampling-aggregation methods, such as self-consistency, which improve answer stability through multiple reasoning paths, LLM-based multi-agent approaches further introduce explicit mechanisms for interaction, critique, and revision across agents\citep{wang2023selfconsistency}. Multi-Agent Debate enables multiple LLMs to independently propose answers and critique one another over multiple rounds, thereby improving accuracy in mathematical reasoning, strategic reasoning, and factual question-answering tasks\citep{du2023improving}. Subsequently, ChatEval applied multi-agent debate to text evaluation in LLM-as-a-judge scenarios\citep{chan2024chateval}, while ReConcile enhanced consensus quality in complex reasoning through round-table discussions among heterogeneous LLMs and confidence-weighted voting\citep{chen2024reconcile}. In addition, CAMEL demonstrated that role-playing-based multi-agent dialogue can be used for task decomposition and collaborative exploration\citep{li2023camel}. AutoGen\citep{wu2023autogen}, ChatDev\citep{qian2024chatdev}, and MetaGPT\citep{hong2024metagpt} further reveal the application potential of LLM-based multi-agent systems in tool use, software development, and structured workflows. However, these methods typically rely on fully connected communication, manually specified communication links, or fixed collaborative procedures, and therefore lack explicit control over the trade-off between communication cost and reasoning quality.

\subsection{Sparse Communication Topology}
Communication topology is a crucial factor shaping the efficiency, stability, and quality of collective decision-making in multi-agent systems. In traditional multi-agent reinforcement learning, CommNet learns information exchange among agents through differentiable communication\citep{sukhbaatar2016learning}, while subsequent studies further introduce attention mechanisms, targeted message passing, and communication-triggering mechanisms to learn “with whom to communicate” and “when to communicate”\citep{jiang2018learning, das2019tarmac, singh2019learning}. These studies demonstrate that communication links need not always be fully connected; through selective communication, systems can reduce the propagation of redundant information and enhance collaborative efficiency.

In the context of LLM-based multi-agent debate, existing studies have likewise shown that communication topology significantly influences reasoning efficiency and effectiveness. \citet{li2024sparse} systematically compared multiple sparse communication topologies and found that sparse topologies can preserve, or even enhance, reasoning performance while reducing computational cost. Recent work on sparse mixture-of-agents has also explored sparse information flow in multi-agent LLMs from the perspectives of response selection and early stopping mechanisms\citep{li2024smoa}. Unlike the aforementioned studies, which primarily compare fixed topologies or sparsify communication based on candidate responses, this paper focuses on dynamic topology selection: communication edges are no longer determined by predefined graph structures, but are adaptively generated during the reasoning process according to agent confidence, answer divergence, task state, and historical reliability.

\subsection{Consensus and Trust Modeling}
Traditional consensus algorithms in multi-agent systems typically examine weighted averaging of agent states over communication graphs, neighbor interactions, and convergence conditions\citep{degroot1974reaching, jadbabaie2003coordination, olfati2007consensus}. The consensus problem in LLM-based multi-agent reasoning is more intricate, because agent states encompass natural-language reasoning trajectories, discrete answers, and imperfectly calibrated confidence estimates. Existing studies have shown that the confidence of modern neural networks may suffer from systematic calibration issues\citep{guo2017calibration}; in the context of LLMs, although models can express or estimate their own uncertainty to some extent, such self-assessment is not always reliable\citep{kadavath2022language, lin2022teaching}. Therefore, this paper does not directly assume that LLM outputs satisfy the conditions of classical linear systems, but instead models answer updates and trust-weighted aggregation within an abstract confidence-vector space, and verifies the effectiveness of the dynamic sparse consensus mechanism through experiments with real LLMs.

\section{Problem Definition}

Given a task $x$ to be solved, such as a mathematical problem, a logical reasoning question, or a factual question-answering task, the system comprises $n$ LLM agents $\mathcal{A}={a_1,\ldots,a_n}$. Each agent maintains a state at round $t$:
\begin{equation}
z_i^t = (y_i^t, r_i^t, c_i^t),
\end{equation}
where $y_i^t$ denotes the current answer, $r_i^t$ denotes the natural-language reasoning or explanation, and $c_i^t \in [0,1]$ denotes the self-reported or calibrated confidence.

In fully connected multi-agent debate, the communication graph at round $t$ is $G_t=(V,E_t)$, where $V=\mathcal{A}$ and $E_t={(i,j):i\neq j}$. Thus, each round contains $n(n-1)$ directed messages. This paper considers sparse communication under budget constraints: each agent receives information from at most $k$ neighbors, where $k \ll n$. The objective is to maximize the quality of the final answer under a communication budget $B$, while reducing token consumption and end-to-end latency:
\begin{equation}
\max_{{E_t}{t=1}^{R}} ; \mathbb{E}[Q(\hat{y}, y^*)]
\quad \text{s.t.} \quad \sum{t=1}^{R}|E_t| \leq B,
\end{equation}
where $\hat{y}$ denotes the final consensus answer, $y^*$ denotes the ground-truth answer, and $Q$ denotes a task-quality metric, such as exact match or correctness.

\section{Method}

DySCo comprises four modules: individual initial reasoning, dynamic edge selection, trust-aware state updating, and final consensus aggregation. The overall procedure is presented in Algorithm~\ref{alg:dysco}.

\subsection{Individual Initialization}

At round 0, each agent independently solves the task $x$:
\begin{equation}
(y_i^0, r_i^0, c_i^0) = \mathrm{LLM}i(P{\mathrm{solve}}(x)),
\end{equation}
where $P_{\mathrm{solve}}$ denotes a unified task-solving prompt. The agents may be different sampling instances of the same LLM, or heterogeneous agents instantiated through different models, different system prompts, or distinct role specifications.

\subsection{Dynamic Communication Edge Selection}

At the beginning of round $t$, DySCo computes a communication-value score $s_{ij}^t$ for each potential directed edge $(j \rightarrow i)$, representing the prospective benefit for agent $i$ of receiving information from agent $j$ in the current round:
\begin{equation}
s_{ij}^t = \alpha T_j^t + \beta c_j^t + \gamma D_{ij}^t + \delta H_{ij}(x) - \eta L_j^t.
\end{equation}
Here, $T_j^t$ denotes the historical trust weight of agent $j$, $c_j^t$ denotes its current confidence, $D_{ij}^t$ represents the degree of answer or reasoning divergence between $i$ and $j$, $H_{ij}(x)$ represents the degree of task dependency or role complementarity, and $L_j^t$ denotes the estimated token cost incurred by receiving the message from $j$. The coefficients $\alpha,\beta,\gamma,\delta,\eta$ are hyperparameters.

The divergence score $D_{ij}^t$ can be obtained from answer agreement, the cosine distance between reasoning embeddings, or judgments produced by a critique prompt. For multiple-choice questions, it can be defined as
\begin{equation}
D_{ij}^t = \mathbb{I}[y_i^t \neq y_j^t].
\end{equation}
For open-ended reasoning tasks, sentence-vector distance may be used to approximate reasoning complementarity. The task-dependency term $H_{ij}(x)$ can be specified by role relationships or a task-dependency graph. For example, in code review, a testing agent may have a high dependency weight with respect to a coding agent; in mathematical reasoning, a verification agent may offer high feedback value to a solving agent.

Under the budget constraint, each receiver $i$ selects the $k$ senders with the highest scores:
\begin{equation}
\mathcal{N}i^t = \operatorname{TopK}{j \neq i}(s_{ij}^t, k),
\end{equation}
and sets $E_t={(j,i): j\in\mathcal{N}_i^t}$. Therefore, $|E_t|\leq nk$.

\subsection{Message Compression and Critique Generation}

To further control communication costs, the sender does not directly broadcast the complete reasoning trajectory, but instead generates a structured, compressed message:
\begin{equation}
m_{j\rightarrow i}^t = \mathrm{LLM}j(P{\mathrm{critique}}(x,y_i^t,r_i^t,y_j^t,r_j^t)).
\end{equation}
The message contains three components: the current answer, the single most critical supporting reason, and a potential counterexample or revision suggestion for the receiver’s answer. This structure constrains the propagation of redundant long-form text, reduces the token budget, and preserves information valuable for consensus.

\subsection{Trust-Aware State Update}

Receiver $i$ updates its answer according to its local state and the messages received from its neighbors:
\begin{equation}
(y_i^{t+1}, r_i^{t+1}, c_i^{t+1}) = \mathrm{LLM}i(P{\mathrm{revise}}(x,z_i^t,{m_{j\rightarrow i}^t,T_j^t}_{j\in\mathcal{N}_i^t})).
\end{equation}
The prompt requires the agent to explicitly distinguish among three cases: “retaining the original answer,” “making a partial revision,” and “completely changing the answer,” while also providing the rationale for any change in confidence. To prevent erroneous agents from dominating the discussion, neighbor messages are presented with weights according to $T_j^t$; the opinions of low-trust agents are required to be treated only as counterexamples awaiting verification, rather than as direct evidence.

Trust weights are updated according to historical performance, current consistency, and verifier feedback:
\begin{equation}
T_i^{t+1}=\lambda T_i^t+(1-\lambda)(\rho V_i^t+(1-\rho)A_i^t),
\end{equation}
where $V_i^t$ denotes the score assigned by a lightweight verifier to agent $i$’s current reasoning, $A_i^t$ denotes its degree of agreement with the final or provisional majority consensus, and $\lambda$ controls historical smoothing. If the task admits an executable verifier, such as mathematical answer checking, code unit testing, or retrieval-evidence matching, then $V_i^t$ can be provided by an external tool; otherwise, it can be supplied by an independent judge LLM.

\subsection{Consensus Aggregation and Early Stopping}

After at most $R$ rounds, DySCo adopts trust-weighted aggregation:
\begin{equation}
\hat{y}=\arg\max_y \sum_{i=1}^{n} T_i^R c_i^R \mathbb{I}[y_i^R=y].
\end{equation}
For open-ended answers, a normalizer may first be used to map equivalent responses into the same candidate set before weighted aggregation is performed. To reduce unnecessary communication, DySCo introduces an early-stopping criterion. When the entropy of the weighted answer distribution falls below the threshold $\epsilon$, or when the consensus answer remains unchanged for two consecutive rounds and the increase in average confidence is smaller than a predefined threshold, the system terminates communication.

\begin{algorithm}[t]
\caption{DySCo: Dynamic Sparse Consensus}
\label{alg:dysco}
\begin{algorithmic}[1]
\REQUIRE Task $x$, agents $\mathcal{A}$, rounds $R$, neighbor budget $k$, stopping threshold $\epsilon$
\STATE Each agent independently generates $(y_i^0,r_i^0,c_i^0)$
\STATE Initialize trust weights $T_i^0 \leftarrow 1/n$
\FOR{$t=0$ to $R-1$}
    \FOR{each receiver agent $i$}
        \STATE Compute edge scores $s_{ij}^t$ for all $j\neq i$
        \STATE Select neighbors $\mathcal{N}_i^t \leftarrow \operatorname{TopK}_{j\neq i}(s_{ij}^t,k)$
    \ENDFOR
    \FOR{each selected edge $(j,i)$}
        \STATE Generate compressed critique message $m_{j\rightarrow i}^t$
    \ENDFOR
    \FOR{each agent $i$}
        \STATE Update local state $(y_i^{t+1},r_i^{t+1},c_i^{t+1})$
        \STATE Update trust weight $T_i^{t+1}$
    \ENDFOR
    \IF{weighted consensus entropy $<\epsilon$}
        \STATE \textbf{break}
    \ENDIF
\ENDFOR
\STATE \textbf{return} weighted consensus answer $\hat{y}$
\end{algorithmic}
\end{algorithm}

\section{Complexity and Consensus Analysis}

\paragraph{Communication Complexity.}
In fully connected communication, the number of messages in each round is $n(n-1)$, and the total number of messages is $R n(n-1)$. If the average length of each message is $\bar{\ell}$, then the communication token cost can be approximated as
\begin{equation}
C_{\mathrm{dense}} = O(Rn^2\bar{\ell}).
\end{equation}
DySCo restricts each agent to receiving at most $k$ messages per round; therefore, the total number of messages is at most $Rnk$, and the communication token cost is
\begin{equation}
C_{\mathrm{sparse}} = O(Rnk\bar{\ell}_c),
\end{equation}
where $\bar{\ell}_c$ denotes the length of the compressed critique message. Since $k\ll n$ and $\bar{\ell}_c\leq\bar{\ell}$, DySCo can substantially reduce communication costs.

\paragraph{Abstract Consensus Stability.}
To analyze the consensus behavior of DySCo, we abstract each agent’s discrete answer as a confidence vector $p_i^t\in\Delta^{|\mathcal{Y}|}$ over the candidate answer set $\mathcal{Y}$. If the communication update at round $t$ can be approximated as
\begin{equation}
p_i^{t+1}=w_{ii}^t p_i^t+\sum_{j\in\mathcal{N}i^t}w{ij}^t p_j^t+\xi_i^t,
\end{equation}
where $w_{ij}^t\geq 0$, $\sum_j w_{ij}^t=1$, and $\xi_i^t$ denotes the nonlinear update error of the LLM, then, under the condition that $\xi_i^t$ is ignored or bounded, this process may be viewed as weighted consensus over a time-varying graph.

\textbf{Proposition 1 (Weak Consensus Condition under a Communication Budget).}
If there exists an integer $B>0$ such that the union graph of any consecutive $B$ rounds of communication is strongly connected, and all nonzero weights satisfy $w_{ij}^t\geq \mu>0$, then, when the perturbation terms $\xi_i^t$ are bounded and have zero mean, the confidence vectors of all agents converge in expectation toward a common consensus neighborhood.

\textit{Proof Sketch.}
In the absence of perturbations, this result corresponds to the classical consensus convergence property of products of time-varying stochastic matrices. If DySCo’s Top-$k$ edge selection preserves sufficient cross-cluster connectivity over time, it can satisfy joint strong connectivity. In the presence of bounded perturbations, the system does not necessarily converge to a single point, but rather to a neighborhood determined by the magnitude of the perturbations. Since real LLM updates are not linear systems, this proposition serves only as an abstract interpretation; its ultimate effectiveness must be established through empirical evaluation.

\begin{figure}[!h]
\centering
\includegraphics[width=\linewidth]{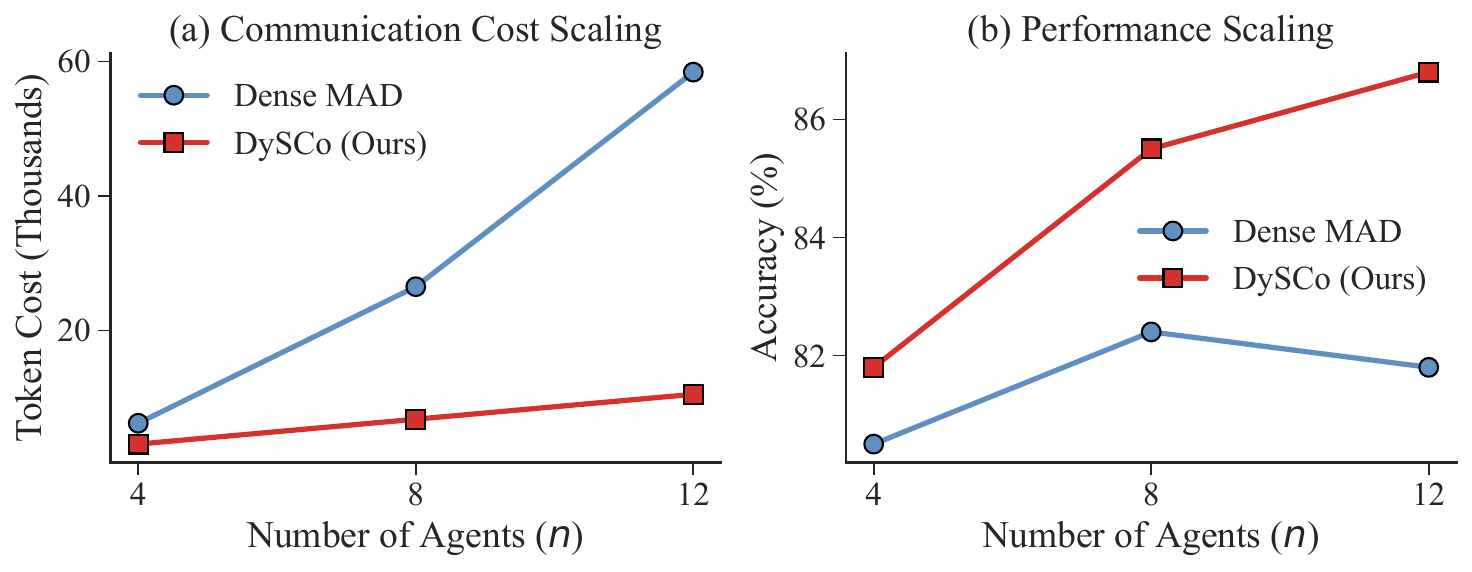}
\caption{Comparison of communication cost and reasoning performance under different agent scales. The left panel shows that, as the number of agents $n$ increases, the token cost of fully connected Multi-Agent Debate rises rapidly, whereas DySCo exhibits a more gradual increase in cost by limiting the number of communication neighbors in each round. The right panel shows that DySCo maintains high accuracy across different agent scales, indicating that dynamic sparse communication can improve scalability while preserving effective multi-agent collaboration.}
\label{fig:scaling}
\end{figure}

\section{Experiment}

\subsection{Experimental Setup}

The experiments cover three categories of tasks: mathematical reasoning, logical and commonsense reasoning, and factual question answering. GSM8K (mathematics), LogiQA (logic), and StrategyQA (commonsense question answering) are selected as the evaluation datasets, respectively. We randomly sample 300 test instances from each dataset for evaluation. The base model is uniformly set to \texttt{gpt-3.5-turbo}, and the number of agents in the multi-agent framework is set to $n=6$ (except in the scalability experiments), with a maximum communication round number of $R=3$. For DySCo and the other sparse topologies, the maximum number of receiving neighbors is set to $k=2$.

We compare DySCo with the following baseline methods: single-agent chain-of-thought reasoning (Single-Agent CoT), self-consistency (Self-Consistency, $n=6$ sampling aggregation), fully connected multi-agent debate (Dense MAD), a fixed ring topology (Static Ring), random selection of $k$ neighbors (Random-$k$), and a sparse topology based solely on historical trust (Trust-Only Sparse). The evaluation metrics include task accuracy (Accuracy), the average total token consumption per question (in thousands), end-to-end average latency (Latency), and consensus entropy (Consensus Entropy, where a lower value indicates smaller disagreement).

\subsection{Main Results}
The main experimental results are shown in Table~\ref{tab:main_results}. The experiments demonstrate that multi-agent methods, including Dense MAD and DySCo, substantially outperform single-agent CoT across all three datasets. However, Dense MAD incurs considerable token consumption and latency. Compared with Dense MAD, DySCo not only reduces token consumption by approximately 70\% and latency by nearly half, but also achieves consistently superior accuracy, reaching 84.3\% on GSM8K, for example. Although fixed topologies, such as Static Ring, and random topologies, such as Random-$k$, effectively reduce overhead, their accuracy gains remain limited because they lose high-value error-correction information. By dynamically matching high-quality neighbors and performing weighted aggregation, DySCo successfully achieves the best trade-off between performance and cost.

\begin{table*}[t]
\centering
\footnotesize
\renewcommand{\arraystretch}{1.16}
\caption{Main performance comparison on GSM8K, LogiQA, and StrategyQA datasets ($n=6, k=2, R=3$). Token usage is measured in thousands (k) per query. Latency is the average end-to-end response time in seconds. Results are reported as mean and standard deviation (subscript) across 3 random seeds. Dark-red annotations in the DySCo row indicate absolute improvements over the strongest dense multi-agent baseline, Dense MAD.}
\label{tab:main_results}
\resizebox{\textwidth}{!}{
\begin{tabular}{L{2.8cm} C{2.2cm} C{2.2cm} C{2.2cm} C{2.2cm} C{2.2cm} C{2.2cm} C{2.2cm}}
\toprule
\multirow{2}{*}{\textbf{Method}} 
& \multicolumn{2}{c}{\textbf{GSM8K}} 
& \multicolumn{2}{c}{\textbf{LogiQA}}
& \multicolumn{2}{c}{\textbf{StrategyQA}}
& \multirow{2}{*}{\textbf{Avg Latency (s)}} \\
\cmidrule(lr){2-3} \cmidrule(lr){4-5} \cmidrule(lr){6-7}
& \textbf{Acc (\%)} $\uparrow$ 
& \textbf{Tokens (k)} $\downarrow$ 
& \textbf{Acc (\%)} $\uparrow$ 
& \textbf{Tokens (k)} $\downarrow$ 
& \textbf{Acc (\%)} $\uparrow$ 
& \textbf{Tokens (k)} $\downarrow$ 
& \\
\midrule
Single-Agent CoT & $72.3_{\pm 0.8}$ & $0.4_{\pm 0.0}$ & $68.1_{\pm 0.9}$ & $0.5_{\pm 0.0}$ & $75.6_{\pm 0.7}$ & $0.3_{\pm 0.0}$ & $3.2_{\pm 0.3}$ \\
Self-Consistency & $78.5_{\pm 0.5}$ & $2.5_{\pm 0.1}$ & $72.4_{\pm 0.7}$ & $3.1_{\pm 0.1}$ & $80.2_{\pm 0.5}$ & $1.9_{\pm 0.1}$ & $4.5_{\pm 0.4}$ \\
Dense MAD & $82.0_{\pm 1.2}$ & $14.8_{\pm 0.8}$ & $76.8_{\pm 1.1}$ & $16.2_{\pm 0.9}$ & $83.5_{\pm 1.0}$ & $12.4_{\pm 0.7}$ & $18.6_{\pm 2.1}$ \\
Static Ring & $80.1_{\pm 1.1}$ & $4.2_{\pm 0.2}$ & $74.3_{\pm 1.2}$ & $4.8_{\pm 0.2}$ & $82.0_{\pm 0.9}$ & $3.6_{\pm 0.1}$ & $9.4_{\pm 0.8}$ \\
Random-$k$ & $80.5_{\pm 1.5}$ & $4.2_{\pm 0.2}$ & $74.5_{\pm 1.4}$ & $4.8_{\pm 0.2}$ & $81.8_{\pm 1.3}$ & $3.6_{\pm 0.1}$ & $9.5_{\pm 0.9}$ \\
Trust-Only Sparse& $81.8_{\pm 0.9}$ & $4.2_{\pm 0.2}$ & $75.6_{\pm 0.8}$ & $4.8_{\pm 0.2}$ & $83.1_{\pm 0.7}$ & $3.6_{\pm 0.1}$ & $9.6_{\pm 0.8}$ \\
\midrule
\rowcolor{ourgray}
\textbf{DySCo (Ours)}
& \metricimp{84.3}{0.4}{\uparrow 2.3}
& \metricimp{4.5}{0.3}{\downarrow 10.3}
& \metricimp{78.5}{0.6}{\uparrow 1.7}
& \metricimp{5.0}{0.3}{\downarrow 11.2}
& \metricimp{85.4}{0.4}{\uparrow 1.9}
& \metricimp{3.8}{0.2}{\downarrow 8.6}
& \metricimp{9.8}{1.0}{\downarrow 8.8} \\
\bottomrule
\end{tabular}
}
\end{table*}

\begin{table*}[!h]
\centering
\begin{minipage}[t]{0.48\textwidth}
\centering
\caption{Ablation study on the GSM8K dataset. Token Delta indicates the percentage change in token usage compared to the full DySCo model.}
\label{tab:ablation}
\resizebox{\linewidth}{!}{
\begin{tabular}{lccc}
\toprule
\textbf{Variant} & \textbf{Acc (\%)} & \textbf{Token Delta} & \textbf{Consensus Entropy} \\
\midrule
\rowcolor{ourgray} \textbf{Full DySCo} & \textbf{84.3} & \textbf{--} & \textbf{0.21} \\
w/o Trust Weight & 81.5 & +1.2\% & 0.45 \\
w/o Diversity Score & 82.1 & -0.5\% & 0.38 \\
w/o Task Dependency & 83.2 & +0.0\% & 0.25 \\
w/o Msg Compression & 84.5 & +82.2\% & 0.20 \\
w/o Early Stopping & 84.3 & +25.4\% & 0.18 \\
\bottomrule
\end{tabular}
}
\end{minipage}
\hfill
\begin{minipage}[t]{0.48\textwidth}
\centering
\caption{Scalability performance on GSM8K with varying number of agents ($n$). Metric $k$ is dynamically scaled as $\lfloor n/3 \rfloor$ for DySCo.}
\label{tab:scalability}
\resizebox{\linewidth}{!}{
\begin{tabular}{lccccc}
\toprule
\multirow{2}{*}{\textbf{Agents ($n$)}} 
& \multirow{2}{*}{\textbf{DySCo $k$}}
& \multicolumn{2}{c}{\textbf{Dense MAD}}
& \multicolumn{2}{c}{\textbf{DySCo (Ours)}} \\
\cmidrule(lr){3-4} \cmidrule(lr){5-6}
& & \textbf{Acc (\%)} & \textbf{Tokens (k)} & \textbf{Acc (\%)} & \textbf{Tokens (k)} \\
\midrule
$n=4$  & $1$ & 80.5 & 6.2  & 81.8 & 3.1 \\
\rowcolor{ourgray}
$n=6$  & $2$ & 82.0 & 14.8 & \textbf{84.3} & \textbf{4.5} \\
$n=8$  & $2$ & 82.4 & 26.5 & 85.5 & 6.8 \\
$n=12$ & $4$ & 81.8 & 58.4 & 86.8 & 10.5 \\
\bottomrule
\end{tabular}
}
\end{minipage}
\end{table*}

\subsection{Ablation Study}

\paragraph{Module Ablation.} Ablation experiments were conducted on the datasets, and the results are shown in Table~\ref{tab:ablation}. Removing trust awareness (w/o Trust Weight) and removing divergence evaluation (w/o Diversity Score) both lead to a marked decline in accuracy, demonstrating that selecting neighbors who are both reliable and endowed with complementary perspectives is essential for error correction. Removing message compression (w/o Message Compression) yields a slight accuracy improvement of 0.2\%, but causes token consumption to surge by 82.2\%. Removing the early-stopping mechanism (w/o Early Stopping) introduces unnecessary communication rounds, increasing cost without delivering any gain in accuracy.

\paragraph{Scalability with Agent Count.}

The central challenge faced by existing multi-agent frameworks is the explosion of communication cost brought about by scaling. Table~\ref{tab:scalability} presents the performance of different frameworks on GSM8K when the number of agents is $n \in {4, 8, 12}$. As $n$ increases, the token consumption of Dense MAD grows quadratically and sharply, reaching a striking 58.4k when $n=12$; moreover, because erroneous opinions propagate indiscriminately, its accuracy improvement tends to plateau and even decline when $n>8$. In contrast, DySCo maintains linear communication-cost growth of $O(Rnk)$ and, by precisely distilling high-quality opinions, enables system accuracy to rise steadily with agent scale, reaching 86.8\%.

\paragraph{Consensus Dynamics Analysis.}

To investigate the effect of dynamic topology on the consensus-convergence process, we tracked the system’s average consensus entropy and accuracy across different communication rounds, as shown in Table~\ref{tab:dynamics}. At round 0, corresponding to initial independent answers, the system exhibits a relatively high entropy value of 1.54 because no communication has yet occurred. After entering the first communication round, DySCo’s consensus entropy rapidly declines to 0.62, substantially lower than the 0.88 observed for Dense MAD. This indicates that trust-based opinion aggregation in DySCo can quickly eliminate salient erroneous disagreements. By round 2, DySCo has largely reached the convergence threshold and triggers extensive early stopping, effectively suppressing the propagation of noise while achieving a higher final accuracy than Dense MAD at round 3.

\begin{table}[b]
\centering
\caption{Evolution of Consensus Entropy and Accuracy over communication rounds on LogiQA.}
\label{tab:dynamics}
\resizebox{\columnwidth}{!}{
\begin{tabular}{lcccc}
\toprule
\multirow{2}{*}{\textbf{Round ($t$)}} 
& \multicolumn{2}{c}{\textbf{Dense MAD}} 
& \multicolumn{2}{c}{\textbf{DySCo (Ours)}} \\
& \textbf{Entropy $\downarrow$} & \textbf{Acc (\%)}
& \textbf{Entropy $\downarrow$}
& \textbf{Acc (\%)} \\
\midrule
$t=0$ (Init) 
& 1.54 & 72.4 
& \cellcolor{gray!15}1.54 
& \cellcolor{gray!15}72.4 \\

$t=1$ 
& 0.88 & 74.5 
& \cellcolor{gray!15}\textbf{0.62} 
& \cellcolor{gray!15}\textbf{76.8} \\

$t=2$ 
& 0.51 & 76.1 
& \cellcolor{gray!15}\textbf{0.28} 
& \cellcolor{gray!15}\textbf{78.2} \\

$t=3$ 
& 0.40 & 76.8 
& \cellcolor{gray!15}\textbf{0.25} 
& \cellcolor{gray!15}\textbf{78.5} \\
\bottomrule
\end{tabular}
}
\end{table}

\begin{table*}[t]
\centering
\footnotesize
\caption{Hyperparameter sensitivity analysis of DySCo. Except for the hyperparameter shown in the current row, all other settings remain consistent with the main experiments; the gray row indicates the default setting used in the main experiments of this paper. Here, $k$ denotes the neighbor budget, $R$ denotes the maximum number of communication rounds, $\epsilon$ denotes the early-stopping threshold, and $\alpha$, $\beta$, $\gamma$, $\delta$, and $\eta$ denote the historical trust, current confidence, answer divergence, task relevance, and communication-cost penalty terms, respectively. The results show that DySCo maintains stable performance under different hyperparameter settings, indicating that its effectiveness does not depend on any particular set of manually tuned parameters.}
\label{tab:hyperparameter_sensitivity}
\begin{tabular}{lccc}
\toprule
\textbf{Setting} & \textbf{Avg. Acc. (\%)} $\uparrow$ & \textbf{Tokens (k)} $\downarrow$ & \textbf{Entropy} $\downarrow$ \\
\midrule
$k=1$ & 81.4 & 3.21 & 0.34 \\
\rowcolor{ourgray}
$k=2$ & 82.7 & 4.43 & 0.25 \\
$k=3$ & 83.0 & 5.86 & 0.22 \\
$R=2$ & 82.0 & 3.72 & 0.31 \\
\rowcolor{ourgray}
$R=3$ & 82.7 & 4.43 & 0.25 \\
$R=4$ & 82.8 & 5.18 & 0.23 \\
$\epsilon=0.20$ & 82.8 & 4.91 & 0.21 \\
\rowcolor{ourgray}
$\epsilon=0.30$ & 82.7 & 4.43 & 0.25 \\
$\epsilon=0.40$ & 82.3 & 3.96 & 0.32 \\
\rowcolor{ourgray}
$\alpha=0.30$ & 82.7 & 4.43 & 0.25 \\
$\alpha=0.40$ & 82.5 & 4.41 & 0.26 \\
\rowcolor{ourgray}
$\beta=0.20$ & 82.7 & 4.43 & 0.25 \\
$\beta=0.35$ & 82.3 & 4.39 & 0.28 \\
\rowcolor{ourgray}
$\gamma=0.25$ & 82.7 & 4.43 & 0.25 \\
$\gamma=0.40$ & 82.9 & 4.48 & 0.24 \\
\rowcolor{ourgray}
$\delta=0.15$ & 82.7 & 4.43 & 0.25 \\
$\delta=0.30$ & 82.4 & 4.44 & 0.27 \\
\rowcolor{ourgray}
$\eta=0.10$ & 82.7 & 4.43 & 0.25 \\
$\eta=0.25$ & 82.1 & 3.91 & 0.30 \\
\bottomrule
\end{tabular}
\end{table*}

\paragraph{Hyperparameter Analysis.}
Table~\ref{tab:hyperparameter_sensitivity} presents the sensitivity of DySCo to key hyperparameters. Overall, DySCo maintains relatively stable performance under different settings, indicating that its effectiveness does not depend on any particular set of manually tuned parameters. For the neighbor budget $k$, when $k$ increases from 1 to 2, the average accuracy rises from 81.4\% to 82.7\%, while consensus entropy decreases from 0.34 to 0.25, suggesting that a moderate increase in communication neighbors helps agents perform effective error correction. However, when $k$ is further increased to 3, accuracy improves only marginally, whereas token consumption increases substantially, indicating diminishing marginal returns from denser communication. For the maximum number of communication rounds $R$, the default setting of $R=3$ achieves a favorable balance between performance and cost; further increasing it to $R=4$ yields only a slight accuracy gain while introducing higher communication cost. The early-stopping threshold $\epsilon$ reflects the trade-off between quality and efficiency: a smaller threshold can achieve lower consensus entropy but requires more communication, whereas a larger threshold can reduce token consumption at the cost of a slight decrease in accuracy. Finally, when the edge-scoring weights $\alpha,\beta,\gamma,\delta,\eta$ are varied, the changes in accuracy, token consumption, and consensus entropy remain modest. This suggests that DySCo’s performance primarily arises from the joint modeling of historical trust, current confidence, answer divergence, task relevance, and communication-cost penalties, rather than from an excessive reliance on any single scoring term or any fixed set of weights.

\section{Discussion}

The central assumption of DySCo is that communication edges in multi-agent collaboration have non-uniform value. Neighbors that exhibit high confidence, strong historical reliability, and meaningful disagreement with the current agent are usually more valuable communication partners than neighbors that are unreliable or merely provide redundant information. A dynamic sparse topology can therefore not only reduce communication cost, but also potentially mitigate the indiscriminate diffusion of erroneous reasoning.

Nevertheless, the method still faces several risks. First, LLM self-reported confidence may be poorly calibrated, and directly using $c_i^t$ may amplify overconfident errors. A possible remedy is to introduce an external verifier or calibrate confidence based on historical task performance. Second, trust weights may produce an early rich-get-richer effect, causing some agents to be marginalized over time. This issue can be alleviated by incorporating an exploration term into edge selection, such as $\epsilon$-greedy exploration or an upper confidence bound. Third, overly sparse communication may lead the system to form local opinion clusters, resulting in a failure to reach global consensus. Therefore, DySCo needs to maintain long-term graph connectivity, for example by periodically introducing random cross-cluster edges.

\section{Limitation}

The effectiveness of DySCo relies on the premise that the value of a communication edge can be approximately estimated from signals such as historical reliability, current confidence, answer divergence, and task relevance. This assumption is relatively reasonable for mathematical reasoning, logical reasoning, and verifiable question-answering tasks, where candidate answers are usually comparable and some errors can be identified through a verifier, a reference answer, or consistency checks. However, in open-ended generation, value-laden judgment, or tasks with multiple valid solutions, these signals do not always reflect answer quality. For example, high confidence may result from model overconfidence, and large answer divergence may reflect only differences in expression rather than substantive complementarity. Therefore, DySCo should be understood as a reasoning coordination mechanism for reducing communication overhead, rather than as a general mechanism for determining factual correctness.

Another limitation arises from the dynamic update of trust weights. Although trust-aware aggregation can suppress the propagation of low-quality opinions, it may also amplify agents that are accidentally correct or more confidently expressed in early rounds, causing subsequent communication to concentrate around a small number of nodes. This mechanism improves efficiency, but it may reduce the continued visibility of minority reasoning paths. For scenarios that require the preservation of dissenting views, evidence-chain inspection, or accountability, DySCo should be combined with explicit exploration edges, human review, or external verification modules, rather than being used as a standalone final adjudication procedure.

{
    \clearpage
    \bibliography{aaai2026_gws}
}
\appendix

\section{Reproducible Prompt Templates}

To facilitate reproducibility, we provide the core prompt templates used by DySCo. 
The templates are written in a model-agnostic form and can be instantiated with different LLM backbones. 
In all experiments, the prompts should be fixed across methods except for communication-specific fields required by each baseline.

\paragraph{Solve Prompt.}
This prompt is used in the initialization stage, where each agent independently solves the task without observing other agents' outputs.

\begin{promptblock}
You are an independent reasoning agent.

Task:
{question}

Instructions:
1. Solve the task independently.
2. Do not refer to other agents or assume that other agents exist.
3. Provide a concise but sufficient reasoning process.
4. If the task is mathematical, show the key intermediate steps.
5. If the task is multiple-choice, select exactly one option.
6. If the task is open-ended, provide the shortest answer that is still complete.
7. Estimate your confidence in the final answer as a number between 0 and 1.

Output format:
Final Answer: <your final answer>
Key Reasoning: <the key reasoning steps>
Confidence: <a number between 0 and 1>
\end{promptblock}

\paragraph{Critique Prompt.}
This prompt is used when an agent sends compressed feedback to a selected neighbor. 
The goal is not to repeat the full reasoning, but to provide a short critique that may help the receiver revise its answer.

\begin{promptblock}
You are a critique agent in a multi-agent reasoning system.

Original Task:
{question}

Your Current Answer:
{sender_answer}

Your Reasoning:
{sender_reasoning}

Neighbor's Current Answer:
{receiver_answer}

Neighbor's Reasoning:
{receiver_reasoning}

Your trust weight in the current round:
{sender_trust_weight}

Instructions:
1. Compare your answer with the neighbor's answer.
2. State whether you agree, partially agree, or disagree.
3. Identify the most important reason supporting your judgment.
4. Point out at most one possible error, missing step, or counterexample
   in the neighbor's reasoning.
5. Do not repeat your full reasoning.
6. Keep the feedback concise and actionable.
7. Use no more than 120 words.

Output format:
Agreement: <agree / partially agree / disagree>
Key Reason: <one concise reason>
Possible Issue: <one possible error or counterexample, or "None">
Suggested Revision: <what the neighbor should reconsider>
\end{promptblock}

\paragraph{Revise Prompt.}
This prompt is used when an agent updates its answer after receiving sparse feedback from selected neighbors. 
Neighbor messages are accompanied by trust weights, so the agent should treat high-trust and low-trust feedback differently.

\begin{promptblock}
You are a reasoning agent revising your answer after receiving feedback
from selected neighboring agents.

Original Task:
{question}

Your Previous Answer:
{current_answer}

Your Previous Reasoning:
{current_reasoning}

Your Previous Confidence:
{current_confidence}

Neighbor Feedback Messages:
{neighbor_feedback_list}

Each feedback message contains:
- neighbor id
- neighbor trust weight
- neighbor answer
- critique message

Instructions:
1. Re-examine your previous answer using the received feedback.
2. Give more consideration to feedback from agents with higher trust weights.
3. Do not automatically follow the majority opinion.
4. Revise your answer only if the feedback exposes a real reasoning error,
   missing evidence, or stronger alternative solution.
5. If you keep your original answer, explain why the feedback is insufficient.
6. If you revise your answer, explain which feedback message was most useful.
7. Output a new confidence score between 0 and 1.

Output format:
Decision: <keep / revise>
New Final Answer: <your updated final answer>
Revision Reason: <why you kept or changed the answer>
Adopted Feedback: <ids of the main feedback messages used, or "None">
New Confidence: <a number between 0 and 1>
\end{promptblock}

\paragraph{Consensus Normalization Prompt.}
For open-ended tasks where exact matching is difficult, we use an optional normalization prompt before trust-weighted voting. 
This step groups semantically equivalent answers without changing their original meaning.

\begin{promptblock}
You are an answer normalization judge.

Original Task:
{question}

Candidate Answers:
{candidate_answers}

Instructions:
1. Group candidate answers that are semantically equivalent.
2. Do not judge which answer is correct unless equivalence requires it.
3. Preserve important distinctions between different numerical values,
   options, entities, or factual claims.
4. Return a canonical answer for each equivalence group.

Output format:
Group 1:
Canonical Answer: <canonical form>
Equivalent Answers: <list of answer ids>

Group 2:
Canonical Answer: <canonical form>
Equivalent Answers: <list of answer ids>
\end{promptblock}
\end{document}